\documentclass[aps,prl,preprint,groupedaddress,showpacs]{revtex
4}
\usepackage{epsfig}
\usepackage{latexsym}

\begin{document}     

\title{Enhanced Thermoelectric Power in Dual-Gated Bilayer 
Graphene } 
\author{Chang-Ran Wang$^{1,\dagger}$, Wen-Sen Lu$^{1,\dagger}$, 
Lei Hao$^1$, Wei-Li Lee$^1$*, Ting-Kuo Lee$^1$, Feng Lin$^2$, 
I-Chun Cheng$^2$ and Jian-Zhang Chen$^3$}      
\affiliation{\textit{$^1$Institute of Physics, Academia Sinica, 
Nankang, Taipei, Taiwan \\$^2$Department of Electrical 
Engineering and Graduate Institute of Photonics and 
Optoelectronics, National Taiwan University, Taipei, Taiwan 
\\$^3$Institute of Applied Mechanics, National Taiwan 
University, Taipei, Taiwan}}

\date{\today}      

\begin{abstract}
Thermoelectric power of a material, typically governed by its 
band structure and carrier density, can be varied by chemical 
doping that is often restricted by solubility of the dopant. 
Materials showing large thermoelectric power are useful for 
many industrial applications, such as the heat-to-electricity 
conversion and the thermoelectric cooling device. Here we show 
a full electric field tuning of thermoelectric power in a 
dual-gated bilayer graphene device resulting from the opening 
of a band-gap by applying a perpendicular electric field on 
bilayer graphene. We uncover a large enhancement in 
thermoelectric power at low temperature, which may open up a 
new possibility in low temperature thermoelectric application 
using graphene-based device.   
               
\end{abstract}
\maketitle
Bilayer graphene (BLG) comprises two monolayers of graphene 
stacked as in bulk graphite exhibiting unusual massive chiral 
fermionic excitations \cite{chiral,martin,barlas}. Its 
conduction band and valence band touch at charge neutral point 
(CNP) resembling a zero-gap semiconductor with an effective 
mass $\rm m^*\approx 0.054m_e$. Many interests reside on the 
band gap engineering in BLG by breaking its inversion symmetry, 
which is of particular importance for semiconductor device 
application. Ohta $\it{et. al.}$ \cite{ohta} demonstrated the 
band gap opening by chemical doping of potassium onto the upper 
layer of a BLG placed on a silicon carbide substrate using 
angle-resolved photoemission spectroscopy. Soon later, Castro 
$\it{et. al.}$ \cite{castro} showed the band gap tuning by 
electric field effect in a BLG device on $\rm SiO_2$(300 nm)/Si 
substrate with $\rm NH_3$ adsorption on the upper layer to 
further enhance the layer asymmetry. More recently, a 
dual-gated geometry for band gap engineering was realized in 
BLG devices \cite{zhang, miyazaki, pablo}, where a BLG was 
sandwiched by top gate and bottom gate providing a full 
electric field control over its band structure. However, 
careful treatment for the top gate dielectric turns out to be 
crucial in order to avoid the downfall of mobility in BLG 
\cite{lin,tutuc}.

In this report, we utilize high quality micro-crystals of 
hexagonal boron nitride (hBN) as top gate dielectric, which was 
shown to impose less trapped charges and strain on graphene 
\cite{dean}. The dual-gated BLG devices we fabricated 
\cite{SOM} show carrier mobility ($\rm\sim$ 2,000-3,000 $\rm 
cm^2$/V-sec) comparable to the ones without top-gate/hBN. We 
demonstrate a full electric-field tuning of thermoelectric 
power (TEP) in bilayer graphene devices, which has been 
predicted theoretically \cite{hao} and originates from the 
opening of a band gap via the application of an out-of-plane 
electrical displacement field $\rm\vec{D}$. We uncover an 
enhancement in TEP due to $\rm\vec{D}$, which grows larger at 
lower temperatures attaining a value comparable to or exceeding 
those of $\rm Bi_2Te_3$-based alloys and sodium cobaltates $\rm 
Na_xCoO_2$ at similar temperature. Our result reveals the 
potential thermoelectric application using graphene-based 
device.

\begin{figure}[ht]
\centerline 
{\epsfig{figure=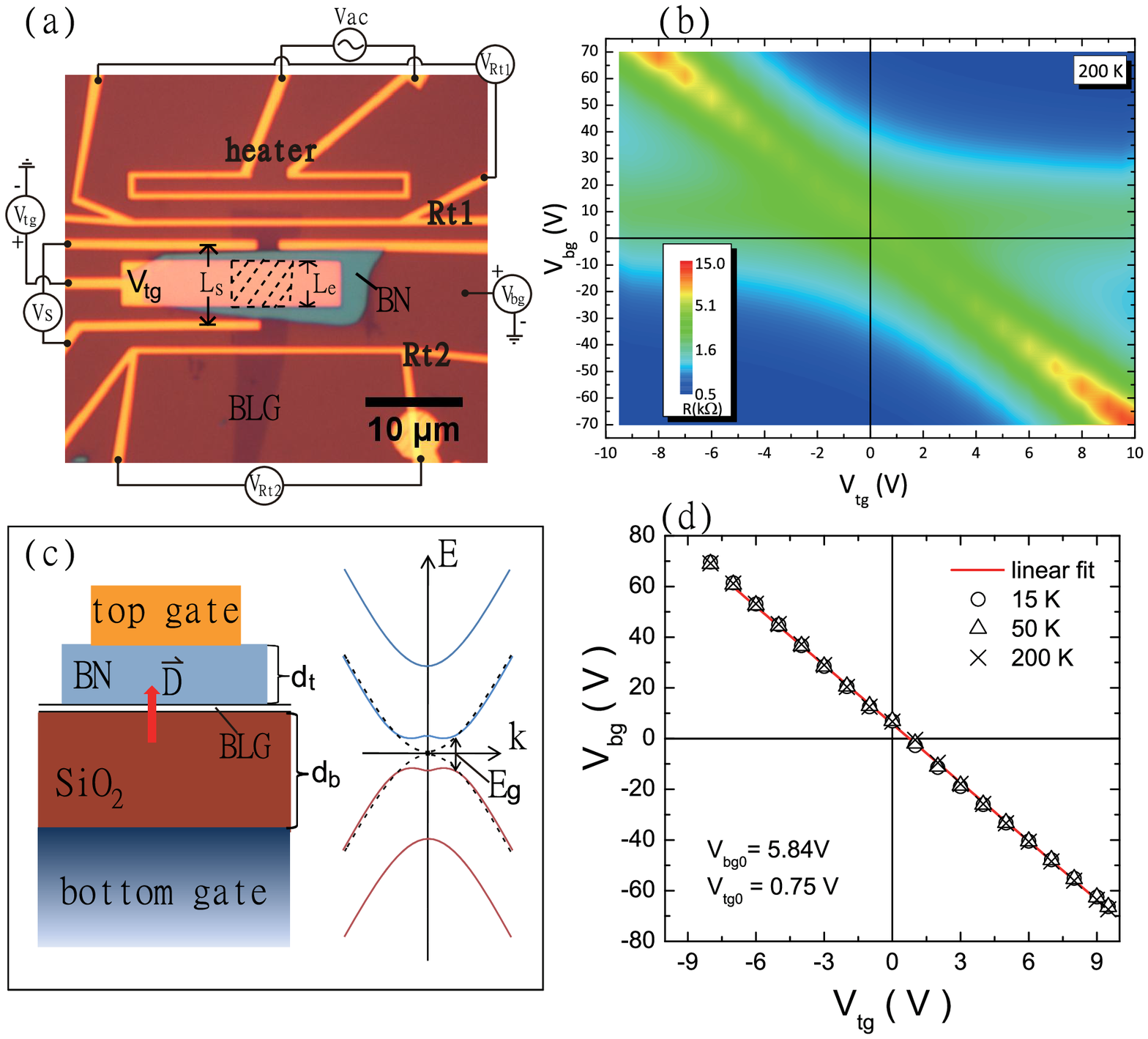,height=3.4in,width=4in,clip
=0}}
\caption {\label{OM_se} (color online) (a) An optical image of 
a dual-gated BLG device with shaded area referring to the 
top-gate effective region. The measurement set-up is 
illustrated with labels in the figure and polarity definitions 
for $\rm V_{tg}$ and $\rm V_{bg}$. (b) A contour plot for the 
resistance (log-scale) as a function of $\rm V_{tg}$ and $\rm 
V_{bg}$ at 200 K. The left panel of (c) illustrates the 
application of displacement field $\rm\vec{D}$ on BLG using 
dual-gated geometry (not to scale). The right panel of (c) 
shows the band gap ($\rm E_g$) opening due to the inversion 
symmetry breaking by $\rm\vec{D}$. (d) The ($\rm V_{tg}$, $\rm 
V_{bg}$) values for resistance peaks in (b) at 15 K, 50 K, and 
200 K. The linear fit shown as red line gives $\rm V_{bg0}$ = 
5.84 V and $\rm V_{tg}$ = 0.75 V.} 
\end{figure}

The device geometry, which comprises a heater, two local 
thermometers (Rt1 and Rt2), three voltages leads and a local 
top gate as shown in Fig. \ref{OM_se}(a), enables the 4-probe 
measurements of resistance and TEP \cite{SOM,TEPsetup,CR}. The 
thickness of the hBN micro-crystal for this particular device 
is $\rm\simeq$ 40 nm determined by an atomic force microscope. 
We note that the top-gate effective region, shown as shaded 
area in Fig. \ref{OM_se}(a), only partially covers the BLG area 
between voltage leads. Therefore, additional geometric factor 
needs to be taken into account in order to extract the sheet 
resistance and TEP for BLG under the top-gate effective region 
\cite{SOM}. For the following discussion, we used the suffix 
e(ne) referring to the top-gate (non)effective region and 
suffix $\rm\Box$ for the sheet resistance.

Figure \ref{OM_se}(b) shows a contour plot for the measured 
resistance R at 200 K for the device shown in Fig. 
\ref{OM_se}(a). The bottom-gate voltage $\rm V_{bg}$ is swept 
up to $\rm\pm$ 70 V, while the top-gate voltage $\rm V_{tg}$ is 
kept at certain value from -10 V to 10V. The high R states 
occur in regions with large $\rm V_{bg}$ and $\rm V_{tg}$ at 
opposite polarity defined in Fig. \ref{OM_se}(a). When plotting 
($\rm V_{tg}$,$\rm V_{bg}$) for the peak position of R (Fig. 
\ref{OM_se}(d)), it is nearly temperature independent and shows 
excellent linearity giving a slope $\rm\alpha\equiv 
dV_{bg}/dV_{tg}=-(\epsilon_{t}d_b)/(\epsilon_{b}d_t)\simeq-7.72
$, where $\rm \epsilon_{b(t)}$ and $\rm d_{b(t)}$ refer to the 
relative bottom(top)-gate dielectric constant and 
bottom(top)-gate dielectric thickness, respectively (Fig. 
\ref{OM_se}(c)). Using $\rm\varepsilon_b$ = 3.9 ($\rm SiO_2$), 
$\rm d_b$ = 300 nm and $\rm d_t$ = 40 nm , we obtained a 
relative dielectric constant for hBN $\rm\varepsilon_t \simeq$ 
4.0 which is in good agreement with the reported value for bulk 
hBN \cite{dean}. The introduction of $\rm V_{bg}$ and  $\rm 
V_{tg}$ capacitively changes the carrier density in BLG and 
hence shifts its chemical potential ($\rm\mu$). For a given 
$\rm V_{tg}$, R attains a peak value whenever $\rm\mu$ is 
shifted back to CNP by tuning $\rm V_{bg}$ while giving a 
finite $\rm\vec{D}$ on BLG that grows in magnitude with $\rm 
V_{tg}$. The total unscreened displacement field $\rm\vec D$ on 
BLG can be calculated using $\rm\vec D 
=[\varepsilon_b(V_{bg}-V_{bg0})/d_b-\varepsilon_t(V_{tg}-V_{tg0
})/d_t]/2$, where $\rm V_{bg0}$ and $\rm V_{tg0}$ equal 5.84 V 
and 0.75 V, respectively, determined from the linear fit shown 
in Fig. \ref{OM_se}(d). Therefore, the large increase in the 
resistance peak is a direct consequence of the band gap opening 
(right panel in Fig. \ref{OM_se}(c)) due to the inversion 
symmetry breaking by $\rm\vec{D}$, which appears to be most 
dramatic in BLG \cite{miyazaki}.

\begin{figure}[ht]
\centerline 
{\epsfig{figure=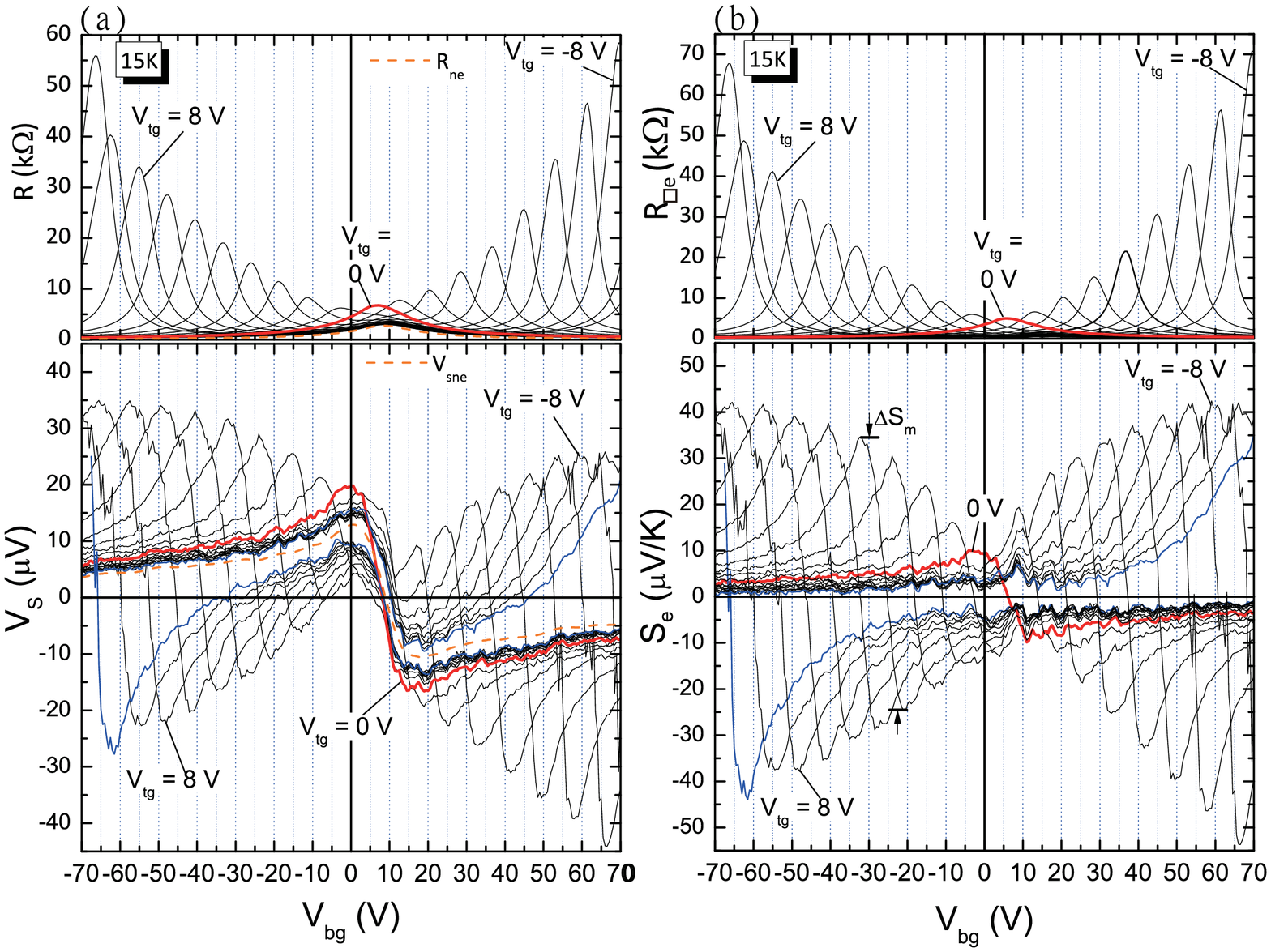,height=3.4in,width=4in,clip=0}}
\caption {\label{rvs} (color online) The upper panel in (a) 
shows R vs. $\rm V_{bg}$ for the dual-gated BLG at 15 K and 
different $\rm V_{tg}$ values ranging from -10 V to 10V. The 
lower panel plots the corresponding thermoelectric signal $\rm 
V_s$ at the same temperature. The dashed orange line represents 
signal contribution from the top-gate noneffective region 
extracted from the blue curves where $\rm V_{tg}= \pm$ 10 V. 
(b) $\rm V_{bg}$ dependence of extracted sheet resistance $\rm 
R_{\Box e}$ (upper panel) and TEP $\rm S_e$ (lower panel) for 
top-gate effective region. The thick red lines are the signals 
at $\rm V_{tg}$ = 0. The definition of $\rm\Delta S_m$ is shown 
in the lower panel of (b).} 
\end{figure}

At $\rm V_{tg}$ = 0 and 15 K, R attains a peak value at $\rm 
V_{bg}\simeq$ 5.84 V shown as thick red line in the upper panel 
of Fig. \ref{rvs}(a). For $\rm V_{tg}\neq 0$, the double-peak 
feature emerges and results from the partial coverage of the 
top-gate region as mentioned earlier. It is then 
straightforward to express R = $\rm R_{e}$+$\rm R_{ne}$, where 
$\rm R_{e(ne)}$ refers to the resistance contribution from 
top-gate (non)effective region. As $\rm V_{tg}$ increases, the 
peak value of $\rm R_{e}$ grows rapidly with its position 
moving to higher $\rm V_{bg}$ value. At $\rm V_{tg}$ = -8 V, 
the increase is nearly 9-fold and tends to grow further at 
higher $\rm V_{tg}$ value. $\rm R_{ne}$, on the other hand, 
shows relatively weak dependence on the $\rm V_{tg}$ and can be 
extracted unambiguously (orange dashed line in Fig. 
\ref{rvs}(a)). The sheet resistance for the top-gate effective 
region $\rm R_{\Box e}=R_eW/L_e$, where W is the width of BLG, 
can then be determined as shown in the upper panel of Fig. 
\ref{rvs}(b).  At $\rm V_{tg}$ = 0, the ratio $\rm 
R_{e}/R\simeq$ 0.58 is close to the length ratio of $\rm 
L_e/L_s\simeq 0.59$ as expected (Fig. \ref{OM_se}(a)).

The thick red curve in the lower panel of Fig. \ref{rvs}(a) 
shows the thermoelectric signal $\rm V_s$ as a function of $\rm 
V_{bg}$ at $\rm V_{tg}$ = 0. $\rm V_s$ is nearly zero at CNP 
and exhibits ambipolar feature where electron-type ($\rm V_s 
<$0) and hole-type ($\rm V_s >$0) carriers can be readily tuned 
by $\rm V_{bg}$. It increases rapidly in magnitude with $\rm 
V_{bg}$ attaining local extremes and falls down at higher $\rm 
V_{bg}$. Similar to the analysis for resistance, we denoted 
$\rm V_s=V_{se}+V_{sne}$, where $\rm V_{se(sne)}$ is the 
thermoelectric signal from the top-gate (non)effective region. 
When turning on $\rm V_{tg}$, $\rm V_{se}$ separates from the 
thick red curve with local extremes occurring at higher $\rm 
V_{bg}$ as $\rm V_{tg}$ increase in magnitude as shown in lower 
panel of Fig. \ref{rvs}(a). The blue curves are $\rm V_s$ 
signals at $\rm V_{tg}=\pm 10 V$ that we used for the 
extraction of the $\rm V_{sne}$ shown as the dashed orange line 
in the lower panel of Fig. \ref{rvs}(b). However, the ratio of 
$\rm V_{se}/V_{s}\sim 0.37$ at $\rm V_{tg}$ = 0 turns out to be 
about 20 \% less than the expected ratio of $\rm L_e/L_s$ 
considering the case of an uniform $\rm (-\nabla T)$ across 
BLG. This implies a smaller temperature gradient $\rm(-\nabla 
T)_e$ under the top-gate effective region due to a likely heat 
shunt through the top-gate(Ti-Au)/hBN. By assuming that TEP at 
zero top-gate voltage is the same throughout the BLG, we can 
then determine $\rm\Delta T_e$ and hence TEP $\rm 
S_e=V_{se}/\Delta T_e$ under the top-gate effective region 
which is shown in the lower panel of Fig.\ref{rvs}(b) 
\cite{SOM}. At 15 K, the difference between local extremes, 
denoted as $\rm\Delta S_m$ (lower panel in Fig.\ref{rvs}(b)), 
equals 20 $\rm\mu$V/K and is enhanced by more than 4 folds to 
$\rm\sim$ 95 $\rm\mu$V/K and 80 $\rm\mu$V/K at $\rm V_{tg}$ = 
-7 V and 8 V, respectively.

\begin{figure}[ht]
\centerline 
{\epsfig{figure=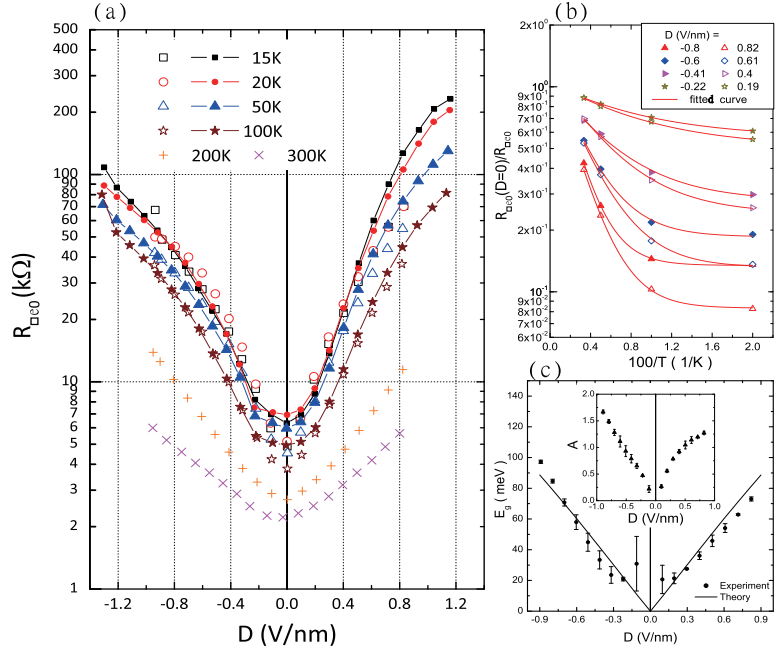,height=3.4in,width=4in,clip=0}}
\caption {\label{gap} (color online) (a) $\rm R_{\Box 
e0}$(log-scale) vs. $\rm\vec{D}$ for the dual-gated BLG at 6 
different temperatures ranging from 15 K to 300 K. $\rm R_{\Box 
e0}$ exponentially grows with $\rm\vec{D}$. (b) The relative 
conductivity $\rm R_{\Box e0}(D=0)/R_{\Box e0}(D)$ vs. 100/T at 
8 different $\rm\vec{D}$ values. The red lines are the fitted 
curves using $\rm R_{\Box e0}(D=0)/\rm R_{\Box e0}(D) = A 
exp(-E_g/2k_BT) + C$. The extracted band gap $\rm E_g$ and A 
parameter as a function of corresponding $\rm\vec{D}$ are shown 
in (c). The solid line is the calculated band gap using 
self-consistent tight-binding model.} 
\end{figure}

The peak values $\rm R_{\Box e0}$ in log-scale are plotted as a 
function of the corresponding $\rm\vec{D}$ values at different 
temperatures shown in Fig. \ref{gap}(a). The solid symbols that 
extend to higher $\rm\vec{D}$ are obtained from the same device 
after cooling down to the base temperature again, which shows 
consistent behavior. $\rm R_{\Box e0}$ increases exponentially 
with $\rm|\vec{D}|$ giving $\rm R_{\Box e0}\simeq$ 230 
k$\Omega$ at $\rm\vec{D}\simeq$ 1.2 V/nm and 15 K, which is 
nearly a 40-fold increase compared to its value at 
$\rm\vec{D}\simeq$ 0 V/nm. However, we notice that the increase 
in $\rm R_{\Box e0}$ drops slightly for $\rm\vec{D}\geq 0.8 
V/nm$. In order to gain further information on the band gap 
$\rm E_g$, we plot the relative conductivity $\rm R_{\Box 
e0}(D=0)/\rm R_{\Box e0}$ as a function of 100/T for eight 
different $\rm\vec{D}$ values such that effects other than 
$\rm\vec{D}$ can be excluded. The data points are then fitted 
in a temperature range of 50 K-300 K using $\rm R_{\Box 
e0}(D=0)/\rm R_{\Box e0} = A exp(-E_g/2k_BT) + C$, where A and 
C are constants independent of T. The extracted $\rm E_g$ vs. 
$\rm\vec{D}$ is shown in Fig. \ref{gap}(c), which is 
practically linear for $\rm\vec{D}\geq$ 0.3 V/nm giving a $\rm 
E_g \simeq$ 100 meV at $\rm\vec{D}\simeq$ -0.9 V/nm close to 
the value given by infrared microspectroscopy \cite{zhang}. The 
data points show good agreement with the calculated band gap 
(solid line in Fig. \ref{gap}(c)) using a self-consistent 
tight-binding model. The fitting function is based on a 
simplified model of a narrow band-gap semiconductor with 
impurity band that may originate from disorder and impurity in 
BLG. The parameter A reflects the $\rm\vec{D}$ dependence of 
carrier mobility $\rm\mu_c\equiv e\tau/m^*$. It turns out to 
increase with $\rm|\vec D|$(inset of Fig. \ref{gap}(c)), where 
the charge re-distribution (screening) and also change in band 
structure may play a role.

\begin{figure}[ht]
\centerline 
{\epsfig{figure=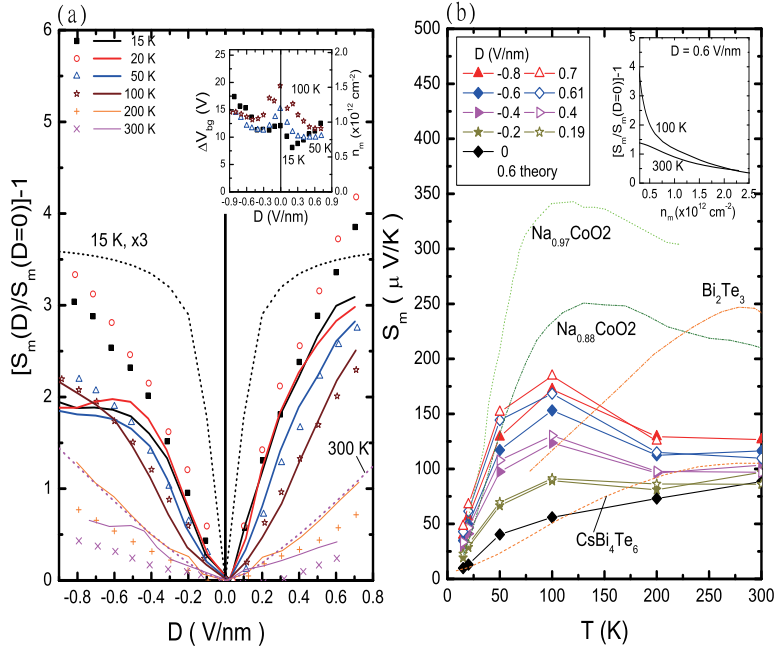,height=3.4in,width=4in,clip=0}}
\caption {\label{SmT} (color online) (a) $\rm 
[S_m(D)/S_m(D=0)]-1$ vs. $\rm\vec{D}$ for the dual-gated BLG at 
6 different temperatures ranging from 15 K to 300 K. The solid 
lines are the relative increase of $\rm[\frac{-T}{R_{\Box 
e}}(\frac{\partial R_{\Box e}}{\partial V_{bg}})]$ obtained 
from the $\rm R_{\Box e}$ data. The dotted and dashed lines are 
the theoretical prediction at 15 K and 300 K, respectively. The 
inset figure plots the width $\rm\Delta V_{bg}$ and its 
corresponding carrier density $\rm n_m$ at which $\rm S_m$ 
occurs. (b) T dependence of $\rm S_m$ for the dual-gated BLG at 
9 different $\rm\vec{D}$ values ranging from -0.8 V/nm to 0.7 
V/nm shown as symbols. The TEP values for $\rm Na_xCoO_2$ 
(x=0.97 and x=0.88), $\rm Bi_2Te_3$ and $\rm CsBi_4Te_6$ are 
shown as dotted lines for comparison. The inset figure shows 
the theoretical calculation of $\rm [S_m(D=0.6 
V/nm)/S_m(D=0)]-1$ vs. $\rm n_m$.}
\end{figure}

We defined $\rm S_m \equiv \Delta S_m/2$ which is the maximum 
value of $\rm S_e$. The increase of $\rm S_m$ relative to its 
value at $\rm \vec{D}=0$, $\rm [S_m(D)/S_m(D=0)]-1$, is shown 
as open symbols in Fig. \ref{SmT}(a). $\rm [S_m(D)/S_m(D=0)]-1$ 
grows larger at lower temperatures and exhibits a minor 
asymmetry with respect to $\rm\vec{D}$. At 20 K, $\rm 
[S_m(D)/S_m(D=0)]-1$ reaches a value of $\rm\sim$ 4.2 at 
$\rm\vec{D}$ = 0.7 V/nm. According to the Mott relation 
\cite{mott}, TEP can be described by 
\begin{equation}
\rm 
S=\frac{\pi^2}{3}\frac{k_B^2}{e}\frac{T}{\sigma}(\frac{\partial
\sigma }{\partial\varepsilon})_{\epsilon=\mu},
\label{eq}
\end{equation}
where $\rm\sigma$ is the electrical conductivity. Using 
$\rm\sigma$=1/$\rm R_{\Box e}$, we can then deduce TEP from the 
sheet resistance data using $\rm 
S_e=\frac{\pi^2}{3}\frac{k_B^2}{e}[\frac{-T}{R_{\Box e}}(\frac{
\partial R_{\Box e}}{\partial V_{bg}})](\frac{\partial 
V_{bg}}{\partial\varepsilon})_{\epsilon=\mu}$ according to Eq. 
\ref{eq}. The increase of the term $\rm[\frac{-T}{R_{\Box 
e}}(\frac{\partial R_{\Box e}}{\partial V_{bg}})]$ with respect 
to its value at $\rm \vec{D}=0$ is shown as the solid lines in 
Fig. \ref{SmT}(a) without any scaling or shifting on them. For 
T $\rm\leq$ 100K and $\rm|\vec{D}|\leq$ 0.3 V/nm, we find 
surprisingly good agreement in the solid lines to the measured 
values of [$\rm S_m(D)/S_m(D=0)$]-1 (symbols). However, the 
solid lines starts to deviate downward from the symbols as 
$\rm|\vec{D}|\geq$ 0.4 V/nm. Based on the Mott relation, the 
discrepancy should reflect the $\rm\vec{D}$ dependence of the 
term $\rm (dV_{bg}/d\epsilon)|_{\epsilon=\mu}$ that is 
proportional to the DOS at $\rm\mu$. We argue that, as 
$\rm|\vec{D}|$ increases, more states near CNP are pushed aside 
forming a band gap that grows wider with $\rm|\vec{D}|$. The 
DOS at the bottom(top) of the conduction(valence) band is then 
expected to increase with increasing $\rm|\vec{D}|$. As shown 
in the inset of Fig.\ref{SmT}(a), $\rm\Delta V_{bg}$, which is 
the width of the local extremes for measured $\rm S_e$, equals 
$\rm\simeq$ 12 V at 15 K corresponding to a carrier density 
$\rm n_m\simeq 1\times 10^{12} cm^{-2}$ at which $\rm S_m$ 
occurs. The gradual increase of $\rm n_m$ at higher 
$\rm|\vec{D}|$ is consistent with the scenario described 
earlier that may also partly account for the $\rm\vec{D}$ 
dependence of A in the inset of Fig. \ref{gap}(c). However, for 
200 K and 300 K, the solid lines deviate upward instead. The 
failure of the Mott relation near CNP at high temperature has 
also been reported previously by several authors 
\cite{CR,korea,shi,hwang}. It was attributed to the violation 
for the criteria of $\rm k_BT/\epsilon_F \ll 1$($\rm\epsilon_F$ 
is the Fermi energy), which becomes more pronounced in a 
cleaner BLG. We note that the enhancement in TEP is mainly 
associated with the change in band curvature due to 
$\rm\vec{D}$ rather than $\rm n_m$ that shows relatively weak 
variation with $\rm\vec{D}$.

The temperature dependence of $\rm S_m$ at different 
$\rm\vec{D}$ values is plotted in Fig. \ref{SmT}(b). It 
exhibits a maximum near 100 K giving a value of $\rm S_m(D=0.7 
V/nm)\simeq 180 \mu$V/K. Below 50 K, we remark that $\rm 
S_m$($\rm\vec D$ = 0.7 V/nm), which tends to increase even 
further at higher $\rm|\vec D|$, is comparable to the reported 
large TEP in sodium cobaltates $\rm Na_xCoO_2$ (x=0.97 and 
0.88) \cite{cobaltate} and $\rm Bi_2Te_3$-based alloys ($\rm 
CsBi_4Te_6$ \cite{CBT} and $\rm Bi_2Te_3$ \cite{BT}) shown as 
dotted lines. Unfortunately, the detailed information on 
thermal conductivity $\rm\kappa$ in dual-gated BLG is absent 
for the determination of thermoelectric figure of merit 
ZT$\rm\equiv S^2\sigma T/\kappa$, where S, $\rm\sigma$, T and 
$\rm\kappa$ are thermoelectric power (TEP), electrical 
conductivity, temperature and thermal conductivity, 
respectively \cite{snyder}. Nevertheless, it was recently 
pointed out that $\rm\kappa$ in encased few-layer graphene can 
be orders of magnitude smaller due to the quenching of flexual 
phonon mode \cite{flexural1,flexural2,lau}, which makes 
dual-gated BLG device a potential candidate for large ZT.

Theoretical calculations including the screening effect are 
based on Kubo's formula of the linear response coefficients 
\cite{hao}. In the clean limit, the predicted relative increase 
of $\rm S_m$ at 15 K and 300 K are shown as dotted and dashed 
lines in Fig. \ref{SmT}(a), respectively, which is more than 
three-fold larger compared to our experimental results. The 
corresponding $\rm n_m$ is, however, an order of magnitude 
smaller than the experimental data. We suspect that the 
presence of charge puddles near CNP can be an important factor, 
which has been revealed from scanning tunneling microscopy in 
exfoliated graphene \cite{puddle,puddle2} and BLG \cite{leroy} 
on $\rm SiO_2$ substrate. The effect of electron-hole puddle 
gives rise to a large charge inhomogeneity $\rm\delta{n}\sim 
10^{12} cm^{-2}$ \cite{puddle,puddle2,tan,yan} near CNP that 
falls in the same order of magnitude as $\rm n_m$ in our device 
. The coexistence of electron-type and hole-type carriers at 
$\rm\mu$ close to CNP is reminiscent of the finite minimum 
conductivity \cite{tan,zhu,sarma,xiao} and also phonon-anomaly 
in bilayer graphene \cite{yan}. This can also cause a 
significant compensation in TEP near CNP as suggested in our 
experiment. We also remark that the relative increase of $\rm 
S_m$ depends critically on $\rm n_m$ as shown in the inset of 
Fig. \ref{SmT}(b) obtained from theoretical calculation for 
$\rm[S_m(D= 0.6 V/nm)/S_m(D=0)]-1$ at different $\rm n_m$. The 
value of $\rm n_m$ inevitably incorporates the impurity doping 
giving rise to larger impurity scattering at higher $\rm n_m$ 
and hence a smaller enhancement in TEP.

In conclusion, we demonstrate the electric-field tunable 
band-gap and TEP in dual-gated BLG device, which may offer a 
new platform for innovative science and engineering. 
Unfortunately, large enhancement in TEP only occurs below 100 K 
in our devices most likely related to the compensation from 
electron-hole puddles near CNP. It gives $\rm S_m$(15K, D = 0.7 
V/nm)=48 $\rm\mu$V/K that is comparable to or exceeding 
existing records for low-T thermoelectric materials. Larger 
enhancement, in principle, can be realized in a cleaner 
dual-gated BLG device at higher $\rm\vec{D}$. With the 
advantage of full electric field control on TEP and also its 
carrier polarity, dual-gated BLG device with a proper design 
can be a promising candidate for low-temperature thermoelectric 
application.

The authors acknowledge the funding support from National 
Science Council in Taiwan (NSC99-2112-M-001-032-MY3) and 
technical support from Core Facility for Nanoscience and 
Nanotechnology at Academia Sinica in Taiwan.


\clearpage

\end{document}